\documentclass[aps,prl,twocolumn,showpacs]{revtex4}

\usepackage{graphicx}
\usepackage{dcolumn}
\usepackage{amsmath}
\usepackage{amssymb}
\usepackage{wasysym}
\usepackage{color}
\begin{document}

\title{On the relationship between charge ordering and the Fermi arcs observed in underdoped high $T_{\rm c}$ superconductors}

\author{N.~Harrison$^1$, S.~E.~Sebastian$^2$
}

\affiliation{$^1$Mail~Stop~E536,~Los~Alamos~National Labs.,Los~Alamos,~NM~ 87545\\
$^2$Cavendish Laboratory, Cambridge University, JJ Thomson Avenue, Cambridge CB3~OHE, U.K
}
\date{\today}

\begin{abstract}
We address the origin of the recently discovered close correspondence between the charge ordering wave vectors and the momentum-space separation between the tips of the Fermi arcs seen in angle-resolved photoemission measurements in underdoped high temperature superconducting cuprates. We calculate the Fermi surface spectral weight for a charge density-wave model, assuming a Fermi surface, charge ordering wave vectors and short correlation lengths similar to those found experimentally. We show that the observation of wavevectors spanning the tips of remnant Fermi surface sections signal Fermi surface reconstruction by charge order, similar to archetypal charge density wave materials, obviating the need to invoke pre-existing Fermi arcs as being unstable to charge ordering. Our findings suggest that charge ordering plays a central role in reconstructing the Fermi surface in underdoped cuprate superconductors.
\end{abstract}
\maketitle

\section{Introduction}
The discovery of charge ordering in a growing number of underdoped high temperature superconducting cuprates raises the possibility of this type of order being universal to the normal state~\cite{wu1,ghiringhelli1,chang1,achkar1,leboeuf1,bakr1,comin1,neto1}. Evidence for charge order is found in x-ray scattering~\cite{comin1,neto1} and Raman spectroscopy~\cite{bakr1} experiments at temperatures as high as the pseudogap onset $T^\ast$. Comparisons of the charge ordering wave vectors found in x-ray scattering and scanning tunneling microscopy (STM) measurements with the Fermi surface spectral weight found in angle-resolved photoemission spectroscopy (ARPES) have revealed an emerging pattern. The charge ordering wave vectors are found to connect the tips of the residual nodal segments of Fermi surface spectral weight within the pseudogap regime termed `Fermi arcs'~\cite{comin1,neto1,sebastian1}. It has been postulated~\cite{comin1,neto1}, based on these observations, that charge ordering is an instability of the Fermi arcs, in which the arcs themselves result from a primary antinodal Fermi surface instability that is distinct from charge ordering~\cite{rice1,lee1}.
\begin{figure}
\centering 
\includegraphics*[width=.48\textwidth]{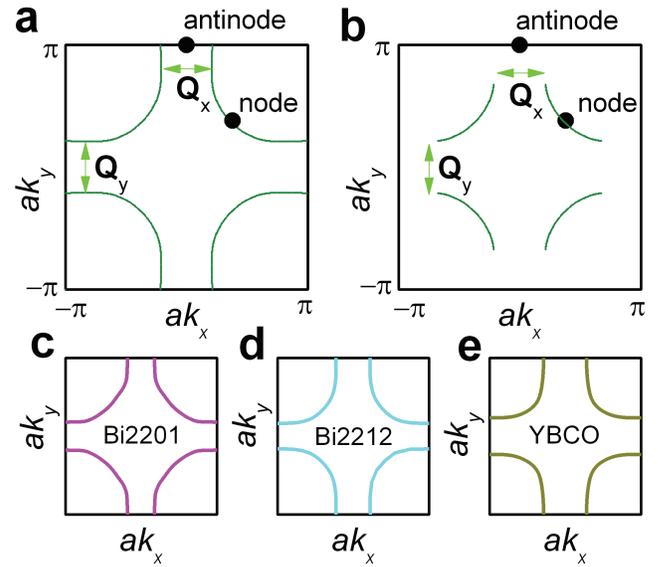}
\caption{({\bf a}), A schematic showing charge ordering wave vectors (${\bf Q}_x$ and ${\bf Q}_y$) approximately nesting the antinodal regions of a large Fermi surface. ({\bf b}), A schematic showing charge ordering wave vectors connecting the 
tips of pre-existing `Fermi arcs,' as has been proposed in Ref.~\cite{comin1,lee1,neto1}.  ({\bf c}, {\bf d}, {\bf e}), Outline of the Fermi surface spectral weight measured by means of ARPES in Bi$_2$Sr$_{2-x}$La$_x$CuO$_{6+\delta}$ (Bi2201)~\cite{meng1}, (Bi,Pb)$_2$Sr$_2$CaCu$_2$O$_{8+\delta}$ (Bi2212)~\cite{kordyuk1} and YBa$_2$Cu$_3$O$_{6.5}$ (YBCO)~\cite{hossain1}. The Fermi surface spectral weight in the antinodal region is observed in all three materials under appropriate experimental conditions.
}
\label{FS}
\end{figure}

In this paper, we show that charge ordering wave vectors spanning the tips of the Fermi arcs are a direct signature of gap formation due to charge density-wave order~\cite{wise1,hanaguri1,shen1,chang1,ghiringhelli1,li1,harrison1,sachdev1,meier1} (see Fig.~\ref{FS}a), rather than being suggestive of pre-existing Fermi arcs that are unstable to charge ordering (see Fig.~\ref{FS}b). On calculating the ARPES spectral weight, considering a scenario in which a Fermi surface instability is caused by charge ordering, we find a correspondence between charge ordering wave vectors and the Fermi arc tips to be a robust feature of the reconstructed Fermi surface. Such a behavior mirrors that seen in several conventional charge density-wave materials~\cite{borisenko1,borisenko2,brouet1,brouet2}. Our calculations therefore suggest that charge ordering is the leading cause of Fermi surface reconstruction in the cuprate materials in which it is observed~\cite{neto1,comin1,sebastian1}.

\section{Fermi surface Model}
We consider a Fermi surface with flat sections in the antinodal regions of the Brillouin zone that are conducive to charge ordering instabilities of characteristic wave vector ${\bf Q}_x$ and ${\bf Q}_y$, as depicted in Fig.~\ref{FS}b. A Fermi surface of this geometry is produced by the tight binding dispersion 
\begin{eqnarray}\label{tightbinding}
\varepsilon_k=-2t[\cos(ak_x)+\cos(ak_y)]+2t^\prime[\cos(ak_x\nonumber\\+ak_y)+\cos(ak_x-ak_y)]-2t^{\prime\prime}[\cos(2ak_x)\nonumber\\+\cos(2ak_y)]-\mu
\end{eqnarray}
on setting $t^\prime/t=$~0.5 and $t^{\prime\prime}/t=$~$-$~0.1, in which we neglect small differences in the lattice dimensions in the $k_x$ and $k_y$ directions. Here, $t$, $t^\prime$ and $t^{\prime\prime}$ are the nearest, next nearest and next next nearest hopping parameters, respectively, while $\mu$ is the chemical potential. These Fermi surface parameters have been chosen based on ARPES measurements on the three cuprates (see Figs.~\ref{FS}c-e) in which charge ordering has recently been detected to high temperatures in x-ray scattering experiments~\cite{comin1,ghiringhelli1,neto1}. All three materials have been shown to exhibit flat portions of the Fermi surface in the antinodal region of the Brillouin zone that are qualitatively similar to Fig.~\ref{FS}b~\cite{meng1,kordyuk1,hossain1}.

\begin{figure*}
\centering 
\includegraphics*[width=1\textwidth]{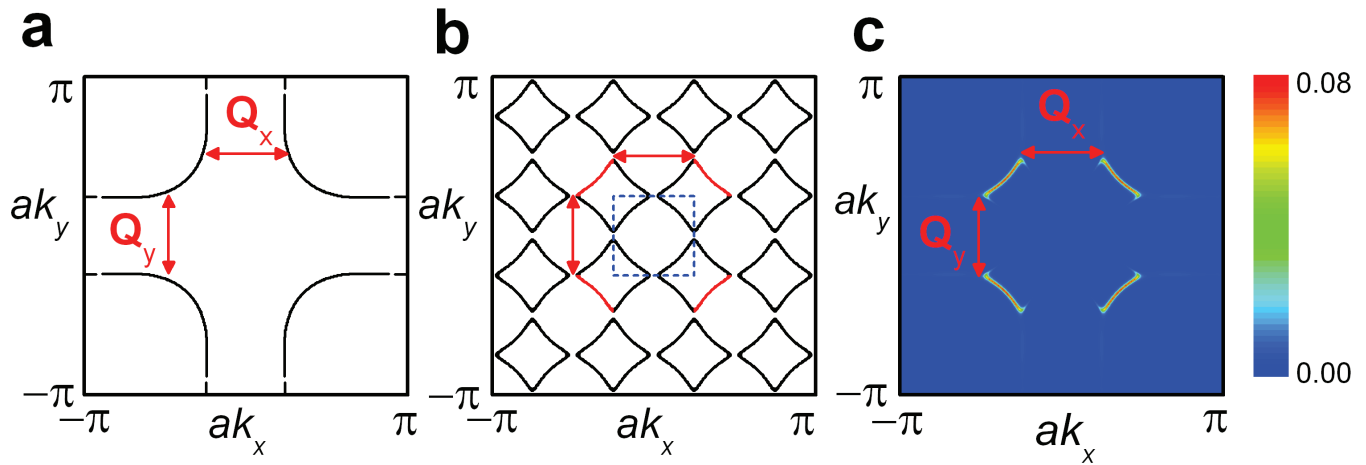}
\caption{({\bf a}), The Fermi surface given by Equation~(\ref{tightbinding}) for an hole doping of $p=$~0.08 relative to the half filled band with the charge ordering wave vectors ${\bf Q}_x$ and ${\bf Q}_y$ shown. Here we use $t^\prime/t=$~0.52 and $t^{\prime\prime}/t=-$0.2. For simplicity we consider a Brillouin zone with a square cross-section. ({\bf b}), The reconstructed Fermi surface produced by Equation (\ref{matrix1}) at the same hole doping, in the repeated Brillouin zone representation. The blue dashed line indicates the reconstructed first Brillouin zone. The sides of the electron pockets giving rise to the Fermi surface spectral weight are indicated in red. ({\bf c}), Contour plot of the Fermi surface spectral weight $A_{\rm bi}({\bf k},\omega)$ calculated using Equation (\ref{spectralweight}), again at the same hole doping, with the charge ordering wave vectors indicated. Each arc corresponds to a different side of the reconstructed pocket.}
\label{reconstructed}
\end{figure*}

\subsection{Long range charge order}
To model the Fermi surface spectral weight seen in ARPES experiments resulting from long range charge order, we consider a two-dimensional density-wave instability that occurs chiefly in the antinodal region of the Brillouin zone, as shown schematically in Fig.~\ref{FS}b. We consider $\varepsilon_k$ states with relative translations of the ordering wave vectors ${\bf Q}_x=(2\pi\delta_x,0)$ and ${\bf Q}_y=(0,2\pi\delta_y)$~\cite{li1,harrison1,norman1,harrison2} to be respectively coupled by $V_x$ and $V_y$, where $\delta_x$ and $\delta_y$ refer to the approximate dimensions of the wave vectors (0.2~$\lesssim\delta_{x,y}\lesssim$~0.3) found in x-ray scattering and STM experiments~\cite{ghiringhelli1,chang1,achkar1,comin1,neto1,wise1,hanaguri1,shen1}. A reconstructed Fermi surface consisting of antinodal gaps at both $a{\bf k}=(\pm\pi,0)$ and $a{\bf k}=(0,\pm\pi)$~\cite{norman1,li1,harrison1} and a nodal electron pocket consistent with quantum oscillation~\cite{doiron1,sebastian1} and Hall effect~\cite{leboeuf2} measurements is produced by a Hamiltonian of the form
\begin{equation}\label{matrix1}
H_{\rm bi}=\left( \begin{array}{cccccccc}
\varepsilon_k & V_x & 0 & V_x & V_y  & 0 & \dots\\
V_x & \varepsilon_{{\bf Q}_x} & V_x & 0 & 0  & V_y & \dots\\
0 & V_x & \varepsilon_{2{\bf Q}_x} &V_x & 0  & 0  & \dots\\
V_x & 0 & V_x & \varepsilon_{3{\bf Q}_x} & 0 & 0 &\dots\\
V_y & 0 & 0 & 0 & \varepsilon_{{\bf Q}_y}  &  V_x &\dots\\
0 & V_y & 0 & 0 & V_x  &  \varepsilon_{{\bf Q}_y+{\bf Q}_x} &\dots\\
\vdots & \vdots & \vdots & \vdots & \vdots & \vdots & \ddots \end{array} \right)
\end{equation}
on adopting suitable forms for $V_x$ and $V_y$~\cite{harrison1,li1,sachdev1}. The diagonal elements $\varepsilon_k$, $\varepsilon_{{\bf Q}x}$, $\varepsilon_{2{\bf Q}x}$~\dots of Equation~(\ref{matrix1}) list the original dispersion in Equation~(\ref{FS}) translated by all possible multiples and combinations of ${\bf Q}_x$ and ${\bf Q}_y$. By approximating $\delta_{x,y}$ with a rational fraction $\delta_x=\delta_y=\frac{m}{n}$, a full Hamiltonian consisting of a $n\times n$ matrix can be constructed, from which $n\times n$ reconstructed electronic bands are obtained upon diagonalization~\cite{harrison1}. 

We neglect bilayer coupling, which, while important for understanding the detailed waveform of quantum oscillations~\cite{sebastian2} and obtaining values of $\delta_x$ and $\delta_y$ closer to those in experiment~\cite{harrison3} in the bilayer cuprates, remains a weak feature in ARPES measurements of the spectral weight deep within the underdoped regime. 

The Fermi surface spectral weight detected in photoemission experiments is obtained by setting $\omega=0$ for the excitation energy in the spectral function~\cite{damascelli1}
\begin{equation}\label{spectralweight}
A_{\rm bi}({\bf k},\omega)=\frac{1}{\pi}{\rm Im}(G_{\rm bi})
\end{equation}
where 
\begin{equation}\label{greensfunction}
G_{\rm bi}({\bf k},\omega)=((\omega+i\Gamma)I-H_{\rm bi})^{-1}_{11}
\end{equation}
is the corresponding Green's function for the Hamiltonian given by Equation~(\ref{matrix1}). The subscript `${11}$' refers to the first diagonal element of the inverted matrix, where $I$ is the identiy matrix and $\Gamma$ represents a simple elastic energy level broadening~\cite{norman1}.

\subsection{Short range charge order}
To simulate the effect of a finite charge order correlation length, as seen in x-ray scattering and STM experiments, we introduce a Gaussian statistical broadening of $\delta$~\cite{comin1} (using $\delta=\delta_x=\delta_y$) in which $\sigma=\frac{a}{2\pi\xi}$ is the standard deviation and $\xi$ is the correlation length. It is convenient, in this case, to consider a simplified Hamiltonian of the form~\cite{norman1} 
\begin{equation}\label{matrix2}
H^\prime_{\rm bi}=\left( \begin{array}{cccccc}
\varepsilon & V_x & V_x & V_y & V_y\\
V_x & \varepsilon_{{\bf Q}_x} & 0 & 0 & 0\\
V_x & 0 & \varepsilon_{-{\bf Q}_x} &0 & 0\\
V_x & 0 & 0 & \varepsilon_{{\bf Q}_y} & 0\\
V_y & 0 & 0 & 0 & \varepsilon_{-{\bf Q}_y}\end{array} \right)
\end{equation}
noting that for period $n$ order the wave vectors $(n-1){\bf Q}_x$ and $(n-1){\bf Q}_y$ are equivalent to $-{\bf Q}_x$ and $-{\bf Q}_y$, respectively. The statistically broadened spectral weight is then obtained by setting $\omega=0$ in
\begin{equation}\label{distribution}
A^\prime_{\rm bi}({\bf k},\omega,\sigma)=\int^\infty_{-\infty}\frac{1}{\pi}{\rm Im}(G^\prime_{\rm bi})\frac{1}{\sigma\sqrt{2\pi}}{\rm e}^{-\frac{(\delta^\prime-\delta)^2}{2\sigma^2}}{\rm d}\delta^\prime
\end{equation}
where 
\begin{equation}\label{greensfunction2}
G^\prime_{\rm bi}({\bf k},\omega)=((\omega+i\Gamma)I-H^\prime_{\rm bi})^{-1}_{11}.
\end{equation}
Below we show that the spectral weight given by Equation~(\ref{distribution}) produces nearly identical results to that given by Equation~(\ref{spectralweight}) on taking the limit $\sigma\rightarrow0$ (i.e. Figs.~\ref{reconstructed}c and \ref{correlation}a), which corresponds to long range order ($\xi\rightarrow\infty$), justifying the use of Equation~(\ref{matrix2}) in calculating the spectral weight.

\section{Results}
Figure~\ref{reconstructed}a  shows the unreconstructed Fermi surface according to Equation~(\ref{tightbinding}) in which $\mu$ has been adjusted to produce a Fermi surface corresponding to a hole filling of $p=$~0.08 (relative to the half-filled band). Figure~\ref{reconstructed}b shows the reconstructed Fermi surface calculated for the same hole doping using $\delta=\frac{1}{4}$ (which lies within the range of values found in x-ray scattering and STM experiments~\cite{ghiringhelli1,chang1,achkar1,comin1,neto1,wise1,hanaguri1,shen1}). We use $V_x=\frac{V_0}{2}(1-\cos ak_y^\prime)$ and $V_y=\frac{V_0}{2}(1-\cos ak_x^\prime)$ (which are chosen to be large only in the antinodal regions of the Brillouin zone where they lead to the opening of a gap at the Fermi surface~\cite{harrison1,note1}) and $V_0/t=0.3$. These parameters are identical to those used in Ref.~\cite{harrison1}. On calculating $A_{\rm bi}({\bf k},\omega)$ in Fig.~\ref{reconstructed}c for the full $n\times n=16\times16$ Hamiltonian given by Equation~(\ref{matrix1}), the Fermi arcs obtained display similarities to those found by Li~{\it et al.}~\cite{li1} and Norman~\cite{norman1}. Here they correspond to one side of the reconstructed electron pocket in Fig.~\ref{reconstructed}b (where they are plotted in red).

\begin{figure}
\centering 
\includegraphics*[width=0.45\textwidth]{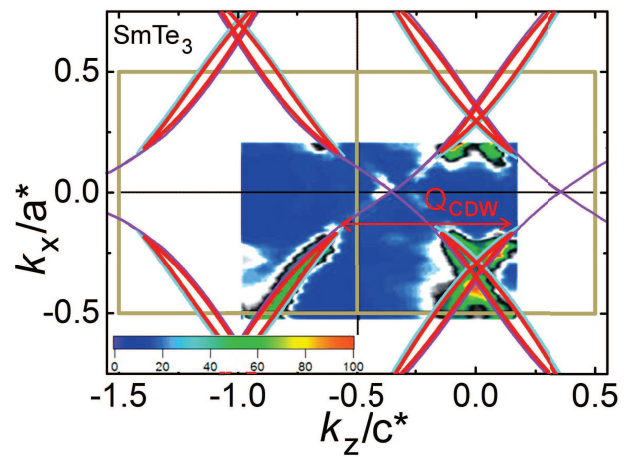}
\caption{ARPES measurements of the SmTe$_3$ Fermi surface from Ref.~\cite{brouet2}. Original 
bands are in violet, bands folded by the charge density-wave are in cyan, obtained from 
band structure calculations. `Fermi arcs' in this case correspond to 
sides of the reconstructed pockets, shown in red. ${\bf Q}$ is seen to span 
the opposite sides of the reconstructed pockets, and consequently 
corresponds to the separation between the arcs. The reduced Brillouin zone is shown in light brown.}
\label{cete3}
\end{figure}

A salient feature of the Fermi surface spectral weight calculated using our charge ordering model in Fig.~\ref{reconstructed}c is that the Fermi arc tips are separated in momentum-space by ${\bf Q}_x$ and ${\bf Q}_y$. This arises from the correspondence of the `Fermi arc'  to one side of the reconstructed electron pocket (plotted in red in Fig.~\ref{reconstructed}b) in this charge ordering model. The spacing between reconstructed electron pockets in the repeated Brillouin zone (shown in Fig.~\ref{reconstructed}b) is consequently given by $2\pi\delta_x={\bf Q}_x$ and $2\pi\delta_y={\bf Q}_y$~\cite{note2},  yielding a separation between Fermi arc tips of ${\bf Q}_x$ and ${\bf Q}_y$. A similar connection between the ends of the arcs in the spectral weight and ${\bf Q}$ vector is found in well-known model charge density-wave systems~\cite{brouet1,brouet2,borisenko1,borisenko2}, as shown for the case of SmTe$_3$ in Fig.~\ref{cete3}.

\begin{figure*}
\centering 
\includegraphics*[width=1\textwidth]{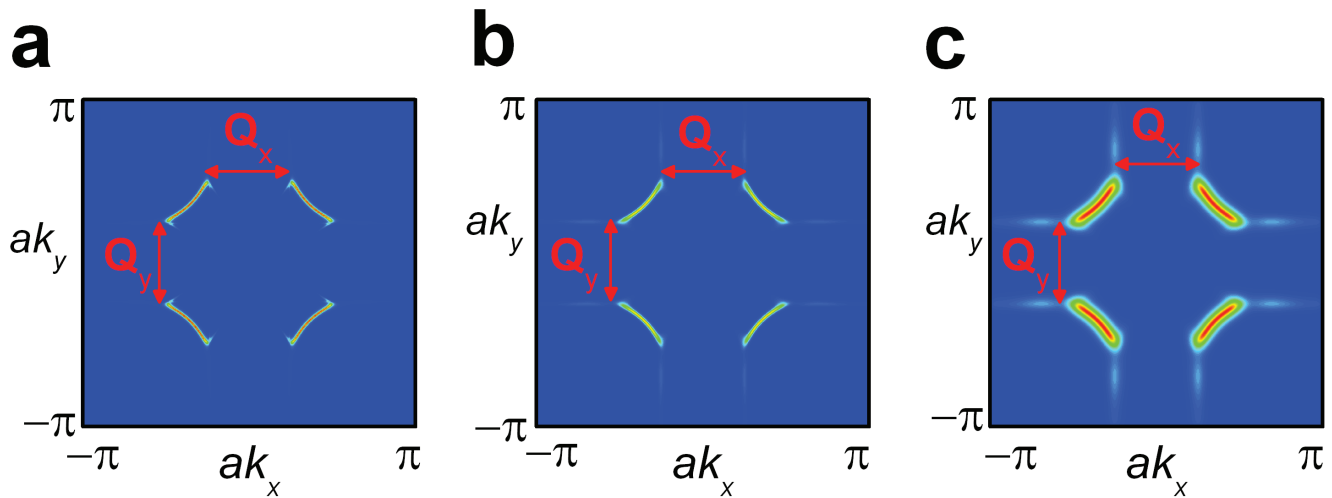}
\caption{({\bf a}), Contour plot of the Fermi surface spectral weight as shown in Fig.~\ref{FS}c in which we have substituted the reduced Hamiltonian $H^\prime_{\rm bi}$ for $H_{\rm bi}$, showing it not to impact the spectral weight. ({\bf b}),  Fermi surface spectral weights calculated using Equation (\ref{distribution}) on including the effect of finite charge ordering correlation length $\xi=$~100. ({\bf c}), Same as {\bf b}, but including a finite ARPES momentum-space resolution. To achieve this, the Fermi surface spectral weight from {\bf b} has been convoluted with a Gaussian of width 0.005~$\times2\pi$.}
\label{correlation}
\end{figure*}


In Fig.~\ref{correlation} we show that the charge ordering wave vectors span the tips of the Fermi arcs on considering a finite correlation length $\xi$. For  $\xi\rightarrow\infty$ in Figs.~\ref{reconstructed}c and \ref{correlation}a, our calculations find band backfolding features at the ends of the Fermi arcs that are associated with the reconstructed electron pocket in Fig.~\ref{reconstructed}b. These features become significantly smeared on including a correlation length $\xi=$~100~\AA~in Fig.~\ref{correlation}b, and more so on simulating the effect of a finite ARPES experimental resolution in Fig.~\ref{correlation}c, potentially explaining their absence in ARPES experimental results obtained in materials with shorter correlation lengths~\cite{hossain1,comin1,neto1,wise1,shen1}. The charge ordering correlation lengths are found at zero magnetic field to be $\xi\approx$~60~\AA~in YBa$_2$Cu$_3$O$_{6+x}$~\cite{ghiringhelli1,chang1} and $\lesssim$~30~\AA~in Bi$_2$Sr$_{2-x}$La$_x$CuO$_{6+\delta}$ and (Bi,Pb)$_2$Sr$_2$CaCu$_2$O$_{8+\delta}$~\cite{comin1,neto1}. Our model also produces a weak spectral weight within the gap in the antinodal regions of the Brillouin zone that is congruent with the unreconstructed Fermi surface. A residual antinodal spectral weight appears to be observed in Bi$_2$Sr$_{2-x}$La$_x$CuO$_{6+\delta}$ ARPES data~\cite{meng1}.

\section{Conclusion}
The recent discoveries of short range or long range charge order extending to high temperatures within the pseudogap regime have led to renewed debate as to the contribution of charge ordering to the formation of the pseudogap and its relevance to superconductivity and competing forms of order~\cite{wu1,ghiringhelli1,chang1,achkar1,leboeuf1,bakr1,sachdev1,lee1,meier1}. One striking aspect of the charge ordering to have emerged is  the close correspondence between its ordering wave vectors and the momentum-space separation of the Fermi arcs seen in ARPES measurements~\cite{comin1,neto1,sebastian1}. Our model shows that such a correspondence is a signature of a Fermi surface reconstructed by charge order, as seen in archetypal charge density-wave materials~\cite{borisenko1,borisenko2,brouet1,brouet2}. 



\section{Acknowledgements}
This work is supported by 
the US Department of Energy BES ``Science at 100 T" grant no. LANLF100, the National Science Foundation and the State of Florida. SES acknowledges funding from the Royal Society and
the European Research Council under the European Union's Seventh
Framework Programme (FP/2007-2013) / ERC Grant Agreement no.
337425-SUPERCONDUCTINGMOTT.

\end{document}